\documentclass[aps,prb,twocolumn,showpacs,superscriptaddress]{revtex4-1}

\usepackage{amsmath,amssymb}
\usepackage{bm}
\usepackage{verbatim}
\usepackage{graphicx}
\usepackage{dcolumn}
\usepackage{amstext}
\usepackage{mathptmx}
\usepackage{verbatim}
\usepackage[colorlinks,citecolor=blue,bookmarks]{hyperref}
\usepackage{times}
\usepackage{xcolor}
\usepackage{color,soul}
\begin{document}


\title{Observation of  Dirac like surface state bands on the top surface of BiSe}


\author{H. Lohani}
\affiliation{Physics Department, Technion-Israel Institute of Technology, Haifa 32000, Israel}
\author{K. Majhi}
\affiliation{Department of Physics, Indian Institute of Science, Bangalore 560012, India}
\author{S.  Gonzalez}
\affiliation{Elettra Sincrotrone Trieste, Strada Statale 14 km 163.5, 34149, Trieste, Italy}
\author{G. Di Santo}
\affiliation{Elettra Sincrotrone Trieste, Strada Statale 14 km 163.5, 34149, Trieste, Italy}
\author{L. Petaccia}
\affiliation{Elettra Sincrotrone Trieste, Strada Statale 14 km 163.5, 34149, Trieste, Italy}
\author{P.S. Anil Kumar}
\affiliation{Department of Physics, Indian Institute of Science, Bangalore 560012, India}
\author{B. R. Sekhar}
\affiliation{Institute of Physics, Sachivalaya Marg, Bhubaneswar 751005, India}
\email[]{sekhar@iopb.res.in}



\date{\today}
\begin{abstract}
 Two quintuple layers (QLs) of strong topological insulator (TI)
Bi$_2$Se$_3$ are coupled by a Bi bilayer in BiSe crystal. We investigated its electronic structure
using angle resolved photoelectron spectroscopy (ARPES)  to study its topological
nature. Dirac like linearly dispersive surface state bands (SSBs) are observed
on the (001) surface of  BiSe and Sb doped (8\%) BiSe, similar to Bi$_2$Se$_3$. Moreover,
 the lower part of the SSBs buries deep in the bulk 
valence band (BVB). Overlap region between the SSBs and BVB is large in Sb doped system
 and the SSBs deviate from the Dirac like linear dispersion in this region.
 These results highlight the role of interlayer 
coupling between the Bi bilayer and the Bi$_2$Se$_3$ QLs.
 Furthermore, we observed a large intensity 
imbalance I(k$_{||}$) $\ne$ I(-k$_{||}$)  in the SSBs located at the positive and negative 
k$_{||}$ directions. This asymmetry pattern  I(k$_{||}$ ) $>$ I(-k$_{||}$)  gradually reverses 
to  I(k$_{||}$) $<$ I(-k$_{||}$)  as the excitation energy scans from low (14eV) to high (34eV) 
value.  However, we did not observe  signal of surface magnetization resulting from the
intensity imbalance in SSBs due to hole-generated uncompensated spin 
accumulation in the photoexcitation process. The main reason for this could be 
the faster relaxation process for photo-hole due to the presence of the Bi bilayer 
between the adjacent Bi$_2$Se$_3$ QLs. The observed  photon energy dependent intensity
 variation could be a signature of the mixing between the spin and the orbit texture of the SSBs.

\end{abstract}


\maketitle


\section{Introduction}

The discovery of spin polarized non-trivial gapless topological surface states 
in bulk insulating materials has been a major breakthrough in the field 
of condensed matter physics\cite{rev,rev1}. Various exotic states of these 
compounds, which are widely known as topological insulators (TIs), are being 
investigated and understood using both experimental and theoretical tools. 
Lots of research work is also being devoted to explore the technologically 
important properties of these materials such as the effect of proximity of TIs 
with superconductors\cite{ab}, correlated and magnetic materials\cite{hong}, 
the spin helical SSBs of TIs to build up some novel devices, especially in the 
area of spintronics\cite{ab1} and quantum computing\cite{maj,maj1}.

Various compounds of the Bi 
chalcogenides family, like Bi$_2$Se$_3$\cite{3dtopo,dirac}, 
Bi$_2$Te$_3$\cite{wrap2} and 
Bi$_{1-x}$Sb$_x$Se$_{1-y}$Te$_y$\cite{bsts,bsts1,sato,hor} exhibit strong TI 
characteristics. BiSe is one of the member of this family which shows a rhombhohedral crystal structure\cite{bise,bise1}. 
This structure consists a Bi$_2$ bilayer between two Bi$_2$Se$_3$ quintuple layers (QLs).
 Bi bilayers has  been recongnised as a quantum spin hall insulators, whereas 
Bi$_2$Se$_3$  is well known strong TI. Therefore, this combined structure of Bi bilayer and 
Bi$_2$Se$_3$ QL {\it i. e.} BiSe is interesting to study  topological properties.
In our previous study we found a clear weak-anti-localization (WAL) cusp in magneto-resistance measurements
in the low filed region. This indicates that BiSe hosts topologically protected surface states\cite{bise}.
The WAL coefficient defined in Hikami-Larkin-Nagaoka (HLN) equation is $\sim$ 0.42 which is close to a value of 0.5.
This is the value expected for a single coherent channel having a $\pi$ Berry phase\cite{hnn}.
DFT predicts two Dirac cones at its side surfaces (100). On the other hand,
SSBs at top surface (001) depends on its stoichiometry. It is Rashba type SSBs
if the top surface contains Bi bilayer while no Dirac like SSBs are found at
Se termination\cite{bise}.
However, it is very hard to directly observe Dirac 
like SSBs by probing the side surfaces of BiSe using ARPES due to 
 non-cleavability of its side surfaces.

In this paper we present our ARPES results on single crystals of BiSe and a Sb 
doped version of it with formula Bi$_{0.92}$Sb$_{0.08}$Se. Dirac like linearly dispersing SSBs 
are observed  on the top surface (001) of BiSe. Furthermore, the Dirac point (DP) is placed deep in 
the bulk valence band (BVB). Sb doping not only increases this overlap between 
the SSBs and BVB but affects the linearity of the SSBs dispersion as well. 
This behavior is quite different from those shown by the known TIs of the Bi 
family Bi$_2$Se$_3$. The difference could be due to the interlayer coupling 
between the Bi bilayer and the Bi$_2$Se$_3$ QL. Interestingly, we noticed 
an appreciable amount of intensity imbalance in the SSBs located at the 
positive and negative k$_{||}$ directions and also it is sensitive to the 
variation in the excitation energy. This possibly indicates a mixing between 
the spin and the orbital texture in the SSBs. These results also highlight the 
role of the Bi bilayer existing between the adjacent Bi$_2$Se$_3$ QLs in this 
material which enhances the relaxation process for the photo-hole generated at 
the SSBs.

\section{Experimental Details}

High quality BiSe single crystals used in this study were synthesized by 
modified Bridgman method. Crystallinity of the samples was confirmed by using 
X-ray diffraction measurements and the elemental composition by energy 
dispersive X-ray (EDX) analysis. Details of other measurements of the physical 
properties can be found elsewhere\cite{bise}. Photoemission measurements were 
performed at the BaDElPh beamline\cite{luca} of the Elettra synchrotron center, 
(Trieste, Italy), equipped with a SPECS Phoibos 150 hemispherical analyser. 
The energy and angular resolutions of the ARPES measurements were set to $\sim$ 20meV 
and 0.3$^\circ$, respectively. The measurements were carried out at 
room temperature under ultra high vacuum conditions with a base pressure of 
$\sim$ 5.0 $\times$ 10$^{-11}$mbar. The single crystalline samples were 
cleaved {\it in-situ} and their orientation were determined by using low 
energy electron diffraction (LEED).

\section{Results and Discussion}

\begin{figure*}
\includegraphics[width=12cm,keepaspectratio]{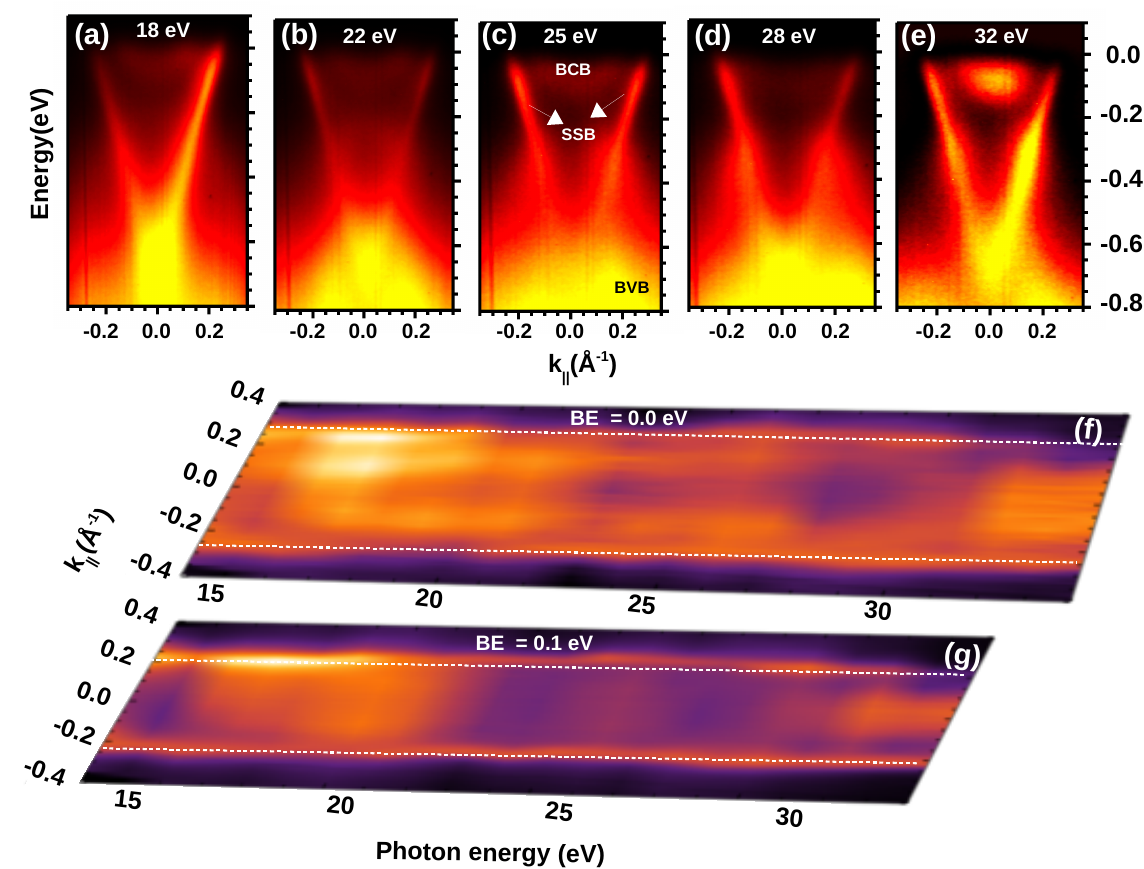}
\caption{\label{fig1}  ARPES images of BiSe taken by using photon energies (p-polarization) 
18eV (a),  22eV (b), 25eV (c), 28eV (d) and 32eV (e). Sample alignment was close to the 
$\Gamma$-M direction. (f) and (g) CEC of BiSe constructed from the MDCs at 
BE = 0.0 and 0.1eV respectively obtained from the ARPES images of photon 
energies (p-polarization) 14eV to 34eV. }
\end{figure*}

Fig.\ref{fig1} shows the results of our ARPES measurements on the BiSe 
crystal by using photons of a wide energy range (14eV to 34eV) performed with 
the aim to map its three dimensional electronic structure. Fig.\ref{fig1}(a)-(e)
depict the high resolution ARPES images of a few selected photon energies (18eV to 32eV).
Two linearly dispersing bands could clearly be seen and between these two bands an intensity patch of parabolic 
shape is also visible close to the Fermi level (E$_f$). Some intensity is
also visible at the higher binding energy (BE) region BE $\simeq$ 0.7eV which 
could correspond to the bulk valence band (BVB). While the linearly dispersive bands remain 
intact as the photon energy varies, a relative change can be 
noticed in the parabolic intensity pattern near the E$_f$.  

To further analyse this point, we constructed constant energy contours (CEC) 
by using the momentum distribution curves (MDC) at BE = 0.0eV (Fig.\ref{fig1}(f)) and 
0.1eV (Fig.\ref{fig1}(g)) from the ARPES images corresponding to photon 
energy of 14eV to 34eV, respectively. Here in Fig.\ref{fig1}(g), the intensity 
pattern shows a constant (white dotted lines are drawn as guide for the eye) 
structure at  k$_{||}$  $\simeq$ $\pm$0.23\AA{}$^{-1}$. This confirms that 
these external bands belong to the surface since they are 
independent of the k$_z$ value associated to  photon energies 14eV to 34eV. 
Furthermore, the variation in the intensity profile between the two linear 
bands indicates the bulk origin of the parabolic band near  E$_f$. 
These bulk and surface originated states are labeled as SSB 
and BVB/bulk conduction band (BCB) in the ARPES image of 
25eV. 

In addition, we found that the SSBs extend quite deep into the BVB. This is 
quite anomalous compared to the known TIs of the Bi chalcogenide family, such 
as Bi$_2$Se$_3$\cite{3dtopo,dirac,bsts1} and BiSbTe$_{1.25}$Se$_{1.75}$\cite{bsts}, 
where the SSBs remains well separated from the BVB and the Dirac point appears
close to the top of the BVB. We have further discussed  this point in the following
paragraph where we compared the ARPES data from BiSe and  Sb doped BiSe.

\begin{figure}
\includegraphics[width=9cm,keepaspectratio]{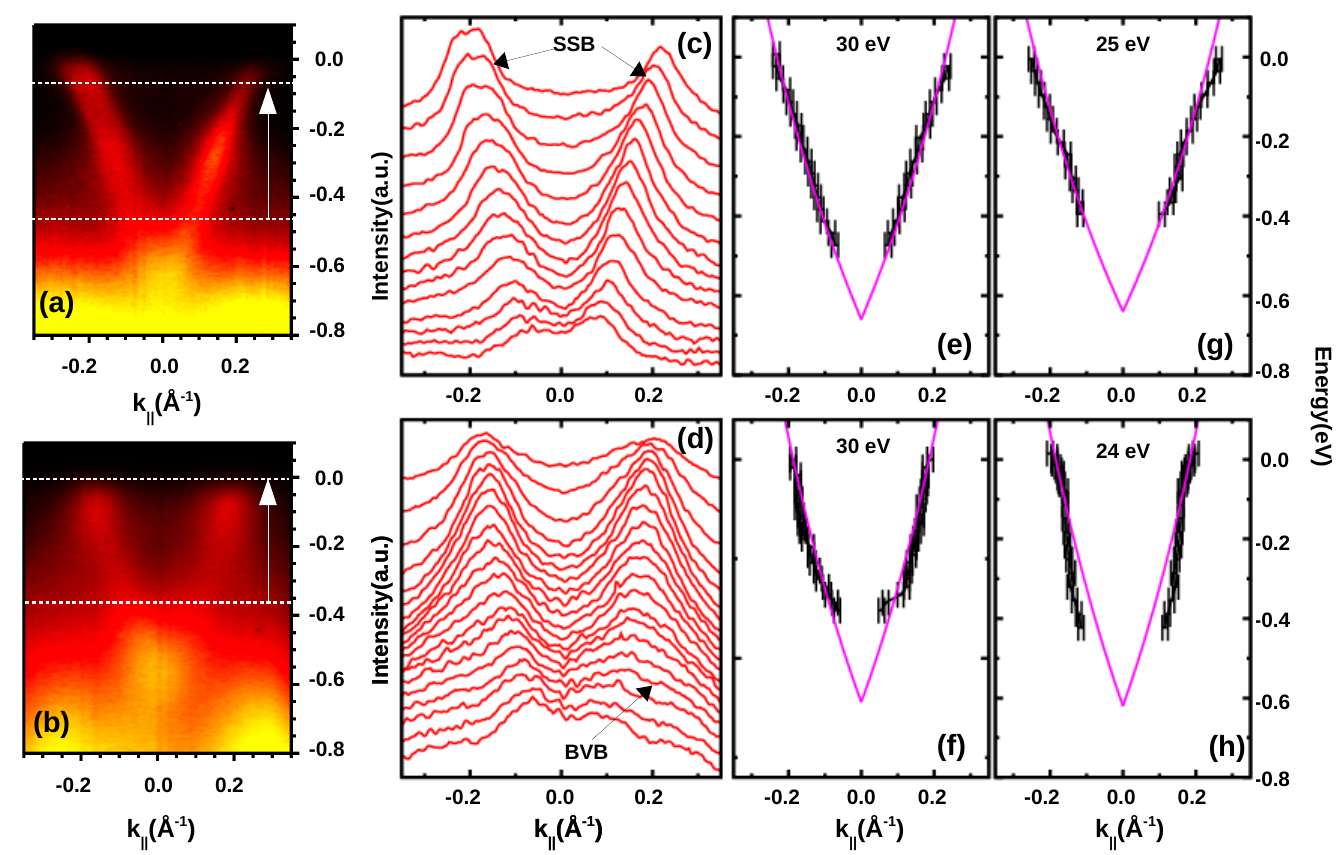}
\caption{\label{fig2}(a) and (b) ARPES images of the $\Gamma$-M oriented 
sample of BiSe and Bi$_{0.92}$Sb$_{0.08}$Se respectively measured at an 
excitation energy of 30eV(p-polarization). (c) and (d) MDC extracted from the ARPES images (a) 
and (b) respectively. Range of MDC extraction is energy range between -0.47eV to -0.08eV for BiSe 
and -0.39eV to 0.0eV  and 0.0eV to -0.6eV for Bi$_{0.92}$Sb$_{0.08}$Se which is also marked by a 
white arrow in the respective images (a) and (b). (e) and (f) dispersion of 
SSBs extracted from the ARPES images (a) and (b) respectively, where magenta 
line represents the calculated SSBs fitted to the experimental data. Similar 
band dispersion of the SSBs obtained from BiSe and Bi$_{0.92}$Sb$_{0.08}$Se
ARPES data at photon energy 25eV and 24eV (Fig.S2 of supp. info. for ARPES images)
 are plotted in (g) and (h) respectively.}
\end{figure}

Fig.\ref{fig2}(a) and (b) show ARPES images of the $\Gamma$-M oriented 
sample of BiSe and Bi$_{0.92}$Sb$_{0.08}$Se, respectively, measured at an 
excitation energy of 30eV. In case of BiSe, the SSBs can be distinguished 
clearly, and in the higher BE region overlap strongly with the BVB states 
positioned at  BE $\simeq$ 0.7eV. Instead, in Bi$_{0.92}$Sb$_{0.08}$Se 
(Fig.\ref{fig2}(b)) BCB states are absent at  E$_f$ and a measurable shift 
towards  E$_f$ is observed in the deeper lying BVB states in comparison to 
those of BiSe. This shifted BVB covers a large portion of the SSBs and they 
are resolved only in the vicinity of  E$_f$. This is also clear from the MDC 
profiles stacked in Fig.\ref{fig2}(c) and (d) extracted from  Fig.\ref{fig2}(a)
and (b) for  energy ranging from -0.47eV to -0.08eV and -0.39eV to 0.0eV, respectively.
 In BiSe the MDC profile at BE close to E$_f$ (Fig.\ref{fig2}(c)) 
shows two well resolved peaks corresponding to the SSBs. As the BE increases 
the two peaks start coming closer to each other and it becomes too difficult to resolve 
them beyond  BE $\simeq$ 0.53eV. On the other hand, in the case of 
Bi$_{0.92}$Sb$_{0.08}$Se (Fig.\ref{fig2}(d)), the MDC  peaks representing the
SSBs are less pronounced. Here also, 
the two SSB peaks become a single peak at higher BE similarly as for BiSe. 
However, a clear hump structure can be noticed  at k$_{||}$ $\simeq$ 0.25\AA{}$^{-1}$ in the MDC profile at 
higher BE which indicates the BVB. In Bi$_{0.92}$Sb$_{0.08}$Se the BVB feature overlaps with the SSBs dispersion,
as inferred from the ARPES images and the MDC plots. 

Furthermore, we extracted the SSBs dispersion from these ARPES images of BiSe and 
Bi$_{0.92}$Sb$_{0.08}$Se by tracking the peak position of the SSB feature 
in the MDC spectra at different BE and results are shown in Fig.\ref{fig2}(e) 
and (f) respectively. These experimental data are also fitted to the following
calculated band dispersion (magenta line)\cite{fu,lu} 
\begin{equation}
\begin{aligned}
&E_{\pm}(\vec{k}) = E_d + k^2/(2m^*)  \pm \sqrt{\nu^{2}_k k^2 + \lambda^2 k^6 cos^2(3\theta)}\\
& \text{here}, \; \quad\nu_k = \nu_0(1 + \alpha k^2)
\end{aligned}
\label{eq1}
\end{equation}
\noindent
where E$_{\pm}$ corresponds to the energy of upper and lower bands, E$_d$ 
is Dirac energy, m$^*$ denotes the effective mass, and 
$\theta$ indicates the azimuthal angle of momentum $\vec{k}$ with respect to 
the x-axis ($\Gamma$-K direction). $\lambda$ is a parameter for the hexagonal 
warping. $\nu_0$ is the Dirac velocity which is modified to $\nu_k$ after 
including a second order correction parameter ($\alpha$) to the Dirac velocity 
in the k $\cdot$ p Hamiltonian.  BiSe data show a reasonably good fit to the calculated bands 
whereas in the case of the Sb doped compound the fitted band slightly deviates 
from the experimental SSBs especially in the higher BE region. In order to 
confirm this difference, the SSB dispersions of BiSe (Fig.\ref{fig2}(g)) and 
Bi$_{0.92}$Sb$_{0.08}$Se (Fig.\ref{fig2}(h)) are compared also at a different 
photon energy. For BiSe the calculated SSBs fit well with the experimental data 
at 25eV photon energy, like the previous case of 30eV photon energy 
(Fig.\ref{fig2}(e)). However, in Bi$_{0.92}$Sb$_{0.08}$Se at excitation energy 
24eV some discrepancy could be seen again at the higher BE region where the 
contribution from the BVB starts to be also significant (see also Fig.S2 of supp. info.). 

Fitting parameters are listed in Table.\ref{table}. We found that the value of $\nu_0$ is $\sim$ 
2.2eV.\AA{} for BiSe which is smaller compared to the value found in other 
compounds of the Bi chalcogenide family such as Bi$_2$Se$_3$ ($\sim$ 
3.2eV.\AA{})\cite{dirac} or BiSbTe$_{1.25}$Se$_{1.75}$\cite{bsts} ($\sim$ 
3.0eV.\AA{}). The table also shows that with Sb doping the m$^*$ of the SSBs 
reduce by $\sim$ 40\% while $\nu_0$ increases by $\sim$ 19\% . This 
result is different from the case of Bi$_2$Se$_3$ thin films where it has been 
found that Sb acts as an electron acceptor for the electronic charge 
segregated at the surface and consequently just pulls the whole electronic 
structure towards the E$_f$ without affecting the $\nu_0$ 
appreciably\cite{thinfilm}. This difference could be due to the different 
interlayer coupling in BiSe; between the Bi bilayer and its adjacent 
Bi$_2$Se$_3$ QLs. The role of Bi bilayer has been thoroughly investigated in 
previous DFT studies and were pointed out the possibility of
various  SSBs in (Bi$_2$)$_n$(Bi$_2$Se$_3$)$_m$\cite{gov,bise}.
In a slab structure (Bi$_2$Se$_3$)$_5$(Bi$_2$)$_2$, where a Bi bilayer is situated on the top of 5 
Bi$_2$Se$_3$ QLs, two   Dirac like SSBs originating from the top and bottom QLs are present
at higher BE compared to original slab of 5 Bi$_2$Se$_3$ QLs. 
 It occurs due to two reasons. One is the Bi bilayer structural relaxation  
to the Bi$_2$Se$_3$ slab and second is a charge transfer from  the
Bi bilayer to the adjacent  Bi$_2$Se$_3$ QL. This  charge transfer
also  creates an internal electric field and that produces a Rashba  splitting
 in the Bi bilayer SSBs which is found at E$_f$. 
In the other case, where the Bi bilayer resides between the
 Bi$_2$Se$_3$ QLs ( (Bi$_2$Se$_3$)$_3$(Bi$_2$)$_2$(Bi$_2$Se$_3$)$_2$ )
  Dirac like SSBs also exist and they  are  localized at the topmost and bottommost QL.
They also resides at higher BE compared to the bare slab of 5 Bi$_2$Se$_3$ QLs.
Thus, in both the situations the SSBs of Bi$_2$Se$_3$ QLs are affected in the presence of Bi bilayer.
This prediction is consistent to current experimental results on BiSe where we found that the DP
at BE $\simeq$ 0.66eV resides at higher BE  in
comparison to  Bi$_2$Se$_3$ ($\simeq$ 0.3eV)\cite{dirac}. Similarly, the SSBs cut
the E$_f$ at larger k$_{||}$ ( k$_f$  $\simeq$ $\pm$ 0.23\AA{}$^{-1}$ ) than in the case of 
Bi$_2$Se$_3$ ( $\simeq$ $\pm$ 0.08\AA{}$^{-1}$ ).

Apart from the different DP and k$_f$ value of the SSBs in BiSe compared
to Bi$_2$Se$_3$, the emergence of the BVB states adjacent to the 
SSBs is also quite peculiar as it influences the evolution of SSBs in doped 
TIs. For example, in case of TlBi(S$_{1-x}$Se$_x$)$_2$ it has been reported 
that while the BVB remains almost intact only the massless Dirac SSBs become 
gradually massive as a function of sulphur doping\cite{newref}. Therefore, 
increment in the overlap region between the BVB and SSBs due to a large 
movement in the BVB towards  E$_f$ shows that the bulk electronic structure 
of BiSe is highly sensitive to the Sb doping. This could again be the result 
of a change in the coupling strength between the Bi bilayer and the Bi$_2$Se$_3$ 
QL due to the dopant. Nevertheless, the existence of SSBs in 
Bi$_{0.92}$Sb$_{0.08}$Se shows that the SSBs are topologically protected 
against this doping. Of course, there are some deviations from the linearity 
in the SSBs dispersion in specific regions where the SSB overlaps with the 
BVB. This reshaping of the SSBs has  recently also been found in Bi$_2$Se$_3$, 
where a non-magnetic impurity produces a kink-like feature in the upper part of 
the SSBs\cite{reson}. 

\begin{table}
\caption{\label{table} Parameters to fit calculated SSB to the experimental SSB 
obtained from the ARPES data of BiSe and Bi$_{0.92}$Sb$_{0.08}$Se.}
\begin{tabular}{ccccc}
\hline
&&BiSe&&\\
\hline
\hline
h$\nu$(eV)&m$^\star$/m&$\nu_0$(eV.\AA{})&$\alpha$(eV.\AA{}$^3$)&E$_d$(eV)\\
\hline
\hline
30&2.1&2.2&2.0&0.66\\
25&1.9&2.0&2.0&0.64\\
\hline
&&Bi$_{0.92}$Sb$_{0.08}$Se&&\\
\hline
\hline
30&1.2&2.5&2.0&0.61\\
24&1.2&2.6&2.0&0.62\\
\hline
\end{tabular}
\end{table}

\begin{figure*}
\includegraphics[width=12cm,keepaspectratio]{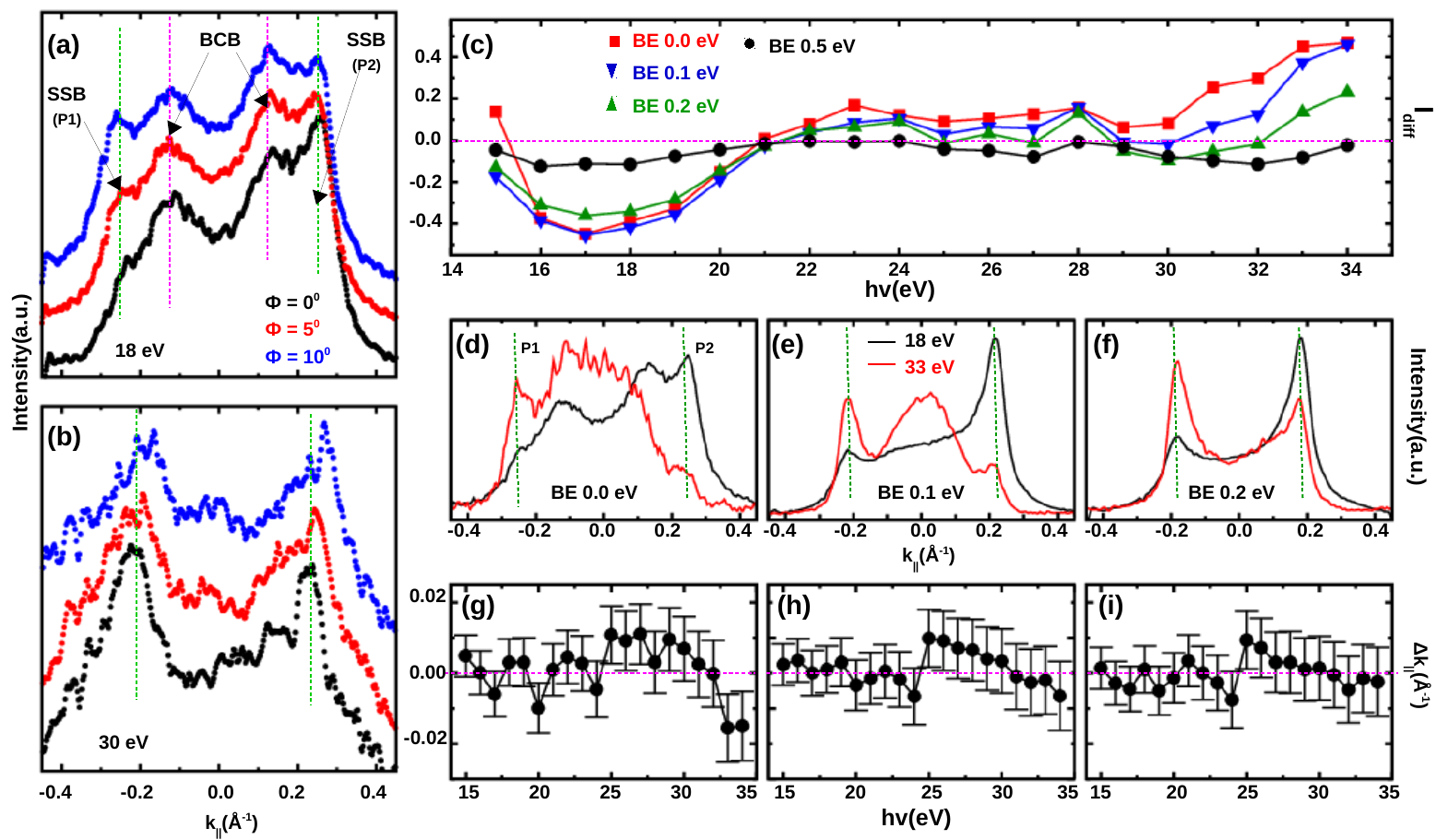}
\caption{\label{fig3} (a) MDC cuts at BE = 0.0 meV of BiSe sample taken at the azimuthal angles 
0$^{\circ}$ (black), 5$^{\circ}$ (red) and 10$^{\circ}$ (blue) 
with respect to $\Gamma$-M direction at photon energy 
18 eV (p-polarization). (b) MDC of BiSe sample taken at the same azimuthal angles 
0$^{\circ}$ (black), 5$^{\circ}$ (red) and 10$^{\circ}$ (blue) at photon energy 
30eV (p-polarization). (c) Intensity ratio (I$_{diff}$ = 
I(P1) - I(P2)) / (I(P1) + I(P2)) between the MDC peaks found in the negative and 
positive k$_{||}$ directions as a function of excitation energy 
 (p-polarization). Red, blue, green and black data correspond to the intensity 
ratio obtained from the MDC profile at BE = 0.0, 0.10, 0.20 and 0.50eV 
respectively. MDC profiles at BE = 0.0eV (d), 0.10eV (e) and 0.20eV (f)) 
extracted from the ARPES images at photon energy (p-polarization) of 18eV 
(black) and 33eV (red). (g), (h) and (i) Shift ($\Delta$k$_{||}$) in the k$_0$ 
value from the  k$_{||}$ = 0.0\AA{}$^{-1}$ for the MDC profile at BE = 0.0, 
0.10 and 0.20eV respectively. Here, k$_0$ refers to the mid point k$_{||}$ value 
between the SSB peaks P1 and P2 of MDC spectra at BE = 0.50eV}
\end{figure*}

Fig.\ref{fig3}(a) shows the MDC at E$_f$ of the BiSe sample taken at the azimuthal angles 
0$^{\circ}$ (black, $\Gamma$-M direction), 5$^{\circ}$ (red) and 10$^{\circ}$ (blue) at photon energy 
18eV. These spectra depict the presence of four peaks in which the outer two 
(green dotted line) belong to the SSBs and the inner two peaks (magenta dotted 
line) correspond to the BCB. It can be noted from this plot that the intensity 
of the SSB peak at positive k$_{||}$ (P2) is larger compared to the peak at 
negative k$_{||}$ (P1).  However, in 
the azimuthal data collected at 30eV (Fig.\ref{fig3}(b)) we found that the 
intensity of the SSB peak P1 increases and becomes comparable to that of the 
SSB peak P2. For further investigation, the intensity ratio 
I$_{diff}$ = (I(P1) - I(P2)) / (I(P1) + I(P2)) from the MDC peaks P1 and P2 is calculated at 
various photon energies. Results are presented in Fig.\ref{fig3}(c) where red, 
blue, green and black data correspond to the intensity ratio obtained from the 
MDC profiles at BE = 0.0, 0.10, 0.20 and 0.50eV, respectively. The intensity 
ratio at  BE = 0.0eV is negative at lower  photon energies (16eV) and it 
gradually increases with the excitation energy, and at higher energies 
(32eV) this ratio becomes positive. The intensity ratio at  BE = 0.10 and 
0.20eV remains almost same at the lower photon energy region (16 to 20eV) 
whereas it decreases at the higher photon energy side (30 to 34eV). To 
illustrate this point, MDC profiles (at BE = 0.0, 0.10 and 0.20eV) extracted 
from the ARPES images at lower and higher photon energies of 18eV (black) and 
33eV (red) are shown in the Fig.\ref{fig3}(d), (e) and (f). It is evident in 
these plots that the SSB peak at negative k$_{||}$ direction (P1) has less 
intensity compared to its counterpart at the positive k$_{||}$ direction (P2) 
at low photon energy and vice versa. On the other side, at higher BE position
 BE = 0.50eV  the intensity ratio does  not  show  any  noticeable  change  as 
 the  photon  energy varies from 14eV to 34eV.
The intensity imbalance remains sensitive to the photon energy variation
 in Sb doped case  as well (see Fig.S3 of supp. info.)

It may appear that this photon energy dependence of the intensity imbalance in 
the k$_x$ - k$_y$ plane could probably be explained in terms of  transition 
matrix elements involved in the photoemission process. In this process 
generally, final states are approximated to plane waves and thereby matrix 
elements mainly depends on the initial states\cite{huf,luca1}. In that case, 
we can expect the intensity ratio to remain the same in Fig.\ref{fig3}(c) for 
all the different BEs because these BEs correspond to various parts of the same 
SSBs. Since this is not the case, the intensity imbalance must be linked to some
other factors. The 
ARPES spectral weight along different k directions can reveal the nature of 
the spin texture of the SSBs in TIs. Previously, it has been reported by 
Jozwiak {\it et al.} that in case of Bi$_2$Se$_3$ the electrons in the SSBs 
at positive and negative k$_{||}$ have the character of up and down spin, 
respectively and this spin helicity interchanges gradually with change in the 
incident light polarization from s to p\cite{soc}. This change is reflected in 
the  spin-resolved photoemission intensity of the SSBs which varies with the polarization.
Furthermore, it has been found that  spin polarization of the  SSBs in Bi$_2$Se$_3$ not
only coupled to crystal momentum but also to the orbital component due to strong
spin-orbit interaction\cite{soc3}. From DFT calculations  Zhu {\it et al.} have revealed the 
wave function of topological surface states ($\psi_{TSS}$) is composed of out-of-plane p$_z$ orbital as well as
in-plane  p$_{x,y}$ orbitals. And  the contribution of different orbitals in the $\psi_{TSS}$
varies at different layers within the topmost 2 QLs in which the $\psi_{TSS}$ is confined.
Spin polarization of the $\psi_{TSS}$ mainly entangles with the in-plane orbital components
and thereby a layer dependent spin-orbital texture is formed. Due to this     
layer dependent $\psi_{TSS}$, photoemission matrix element results  asymmetric intensity
distribution of SSBs in CEC contour (k$_x$,k$_y$) at BE 0.1 eV above the DP. This intensity 
pattern changes to a different pattern as BE varies 0.2 eV above the DP\cite{soc1}. 
Therefore, the asymmetric intensity pattern  I(k$_{||}$) $\ne$ I(-k$_{||}$)  we observed in the SSBs of  BiSe could 
be related to layer dependent  spin-orbital texture.
With this understanding it is difficult to reconcile the photon energy dependency of the 
I$_{diff}$ (Fig.\ref{fig3}(c)) because the orbital character of SSBs is independent to k$_z$
value associated to different photon energies. However, we can understand this
I$_{diff}$ variation in terms of matrix element effects if the final states are not
simple plane waves but have a complex structure, like p or d type states as already pointed out
 by Vidal {\it et al.} in circular dichroism ARPES study of Bi$_2$Se$_3$\cite{soc2}. In this scenario, matrix element which
connects the layer dependent  $\psi_{TSS}$ to different possible final states can lead to photon
energy dependent change in the I$_{diff}$.

Furthermore, in a recent study on pristine and magnetically doped TI 
Bi$_2$Te$_2$Se, Shikin {\it et al.}\cite{luca1} also found an asymmetric 
intensity distribution on the left and right branch of the SSBs and its 
dependency on the photon energy. They proposed that this intensity imbalance 
could develop some surface magnetism under the influence of synchrotron 
radiation due to hole-generated uncompensated spin accumulation. Consequences 
of the in-plane and out-of-plane magnetic moments are a shifted position of 
the DP from k$_{||}$ = 0.0\AA{}$^{-1}$ and a gap opening at the DP position. 
 Both of these signatures are observed prominently in the 
vanadium doped TI Bi$_{1.37}$V$_{0.03}$Sb$_{0.6}$Te$_2$Se. 

In our case, we also observed a very slight asymmetry in the position of the SSBs
 at positive and negative k$_{||}$ direction. To study this point, mid point k$_{||}$ value (k$_0$) between the SSB 
peaks P1 and P2 of MDC profile at BE = 0.50eV is chosen as reference with the 
k$_{||}$ = 0.0\AA{}$^{-1}$ because intensity imbalance is almost absent in this 
MDC spectra (Fig.\ref{fig3}(c)). We estimated the shift ($\Delta$k$_{||}$) in 
the k$_0$ value from the k$_{||}$ = 0.0\AA{}$^{-1}$ for the MDC profile at 
BE = 0.0, 0.10 and 0.20eV for various excitation energies and results are 
depicted in Fig.\ref{fig3}(g), (h) and (i), respectively. Unlike the intensity 
ratio, the $\Delta$k$_{||}$ does not exhibit any significant variation 
(within the error bar limit) with respect to the photon energy.
  In this analysis, the peak positions of P1 and P2 are 
obtained directly from the MDC spectra to determine the k$_0$ because 
asymmetry associated to the MDC spectra hinders a reasonable fitting 
to estimate the peak positions. To calculate the error in the k$_{||}$ 
measurements we used the formula\cite{huf} $\Delta$k$_{||}$ = 
0.512$\sqrt{E_f}$cos($\theta$)$\Delta\theta$ [\AA{}$^{-1}$], where $\theta$ 
represents the photoelectron acceptance angle with respect to the sample normal 
along the analyser slit.  Moreover, the change in the  k$_0$ (Fig.\ref{fig3}(g)) is
 not large enough to quantify the radiation induced magnetization. 
Reason for this could be the absence of magnetic doping and large thermal fluctuations at room 
temperature. On the other hand, the  shift in  k$_0$ is further 
reduced at higher BE (Fig.\ref{fig3}(h) and (i)) despite the significant 
intensity imbalance at these BE (Fig.\ref{fig3}(c)). Therefore, it can also be 
possible that the presence of the Bi bilayer between the adjacent Bi$_2$Se$_3$ 
QLs provides a faster relaxation route to the holes generated at the SSBs and 
thereby the accumulation of uncompensated spin on the SSBs is suppressed. 

\section{Conclusion}

 Our ARPES results on  BiSe  show the presence of Dirac like linearly dispersive SSBs around the 
$\Gamma$ point at its top surface (001). The SSBs are located deep in the BVB 
and with a small (8\%) Sb doping their overlap with the BVB increases 
significantly. The SSBs deviate from the Dirac like linear dispersion in this 
overlap region. This evolution of the SSBs with doping is quite different from 
those shown by other known TIs of the Bi chalcogenide family, like for instance
Bi$_2$Se$_3$, Bi$_2$Te$_3$ and BiSbTe$_{1.25}$Se$_{1.75}$. The reason for 
this difference could be linked to the interlayer coupling between the Bi 
bilayer and Bi$_2$Se$_3$ QL. Another observation of this study is the 
asymmetry in the spectral weight of the SSBs between the positive and negative 
k$_{||}$ directions. This asymmetry gradually switches to the opposite 
direction as the photon energy varies.  This  photon energy dependent intensity
 variation could be a signature of the mixing between the spin and the orbit texture of the SSBs.
We also observed 
that the intensity imbalance in the SSBs  leads an almost zero  shift in the DP 
position in BiSe due to an accumulation of uncompensated spin. 
While the effect is clearly reported for  Bi$_2$Te$_2$Se\cite{luca1}, in BiSe we find 
 a reduced  magnitude of it that can be explained with the presence of  Bi bilayer between 
the adjacent Bi$_2$Se$_3$ QLs of this material.
These  results  have significance in  understanding the spin-orbital texture of the SSBs in BiSe.
These are also helpful for further investigations on  weak TIs. Because theory predicts weak TI
can form by joining  strong TIs via week coupling, like situation in BiSe where a Bi bilayer
couples two QLs of  strong TI Bi$_2$Se$_3$.

{\bf Acknowledgements}:  H.L. is supported in part at Israel Institute of Technology, Technion by 
a  PBC fellowship of the Israel Council for Higher Education.
K.M., P.S.A.K. and B.R.S.
acknowledge the support by Department of Science and Tech-
nology, Government of India for accessing the Elettra Syn-
chrotron.

\end{document}